\RequirePackage{fix-cm}
\documentclass[twocolumn,epjc3]{svjour3}  
\smartqed  
\RequirePackage{graphicx, amssymb, amsmath}
\RequirePackage{latexsym}
\RequirePackage[numbers,sort&compress]{natbib}
\RequirePackage[colorlinks,citecolor=blue,urlcolor=blue,linkcolor=blue]{hyperref}

\protected\def\vt{%
  \ifmmode
    \mskip0.5\thinmuskip
  \else
    \ifhmode
      \kern0.08334em
    \fi
  \fi
} 

\journalname{Eur. Phys. J. C}
\begin{document}

\title{Warp Drives and Martel--Poisson charts
}


\author{Abhishek Chowdhury\thanksref{e1,addr1}
}

\thankstext{e1}{e-mail: achowdhury@iitbbs.ac.in}


\institute{School of Basic Sciences \\Indian Institute of Technology Bhubaneswar, \\Jatni, Khurda, Odisha 752050 India \label{addr1}
}

\date{Received: date / Accepted: date}

\maketitle

\begin{abstract}
We extend the construction of Alcubierre--Nat\'{a}rio class of warp drives to an infinite class of spacetimes with similar properties. This is achieved by utilising the Martel--Poisson charts which closely resembles the Weak Painlev\'{e}--Gullstrand form for various background metrics (Mink, AdS, dS). The highlight of this construction is the non--flat intrinsic metric which in three dimensional spacetimes introduce conical singularities at the origin and in higher dimensions generates non--zero Ricci scalar for the spatial hypersurfaces away from the origin. We analyse the expansion/contraction of space and the (NEC) violations associated with these warp drives and find interesting scalings due to the global imprints of the conical defects. Other properties like tilting of light cones, event horizons and several generalisations are also discussed.   
\end{abstract}

\section{Introduction}
\label{intro}
The idea of ``effective" superluminal travel is of perpetual interest focusing mainly on  ``gedanken--experiments" pushing the foundations of general relativity and communication. Like wormhole geometries \cite{10.1119/1.15620},  warp drive spacetime first proposed by Alcubierre in 1994 \cite{Alcubierre_1994} and then further elaborated upon by Nat\'{a}rio in \cite{Nat_rio_2002} allows for arbitrary large velocities within the framework of general relativity. It appears as a ``bubble" propagating on some spacetime background such that the observers inside the bubble are in an inertial reference frame requiring no external energy sources to accelerate. These ``reaction--less drives" have a severe limitation, sustaining them violates all known energy conditions even for non--relativistic velocities and would probably require exotic matter \cite{Lobo_2004}. Another interesting limitation is the lack of control experienced by an observer inside a superluminal warp bubble which has been addressed by the \textit{Krasnikov tube} \cite{Everett_1997}. 

In this paper we have generalised the construction of Alcubierre--Nat\'{a}rio like warp drive in three dimensional Minkowski spacetime by leveraging the Martel--Poisson (MP) family of charts \cite{Martel_2001} of which the Painlev\'{e}--Gullstrand (PG) coordinates are a special case. This construction has been extended to $\text{AdS}_3$ and dS backgrounds along with their conical defect counterparts. A generic key feature is the appearance of non--flat intrinsic metric with conical singularities which affect the global properties of the spacetimes. Another important realisation comes from the \textit{flow velocities} of the background metrics where the expansion and $(-ve)$ energy density of the warp drive is considerably reduced if the drive travels along the background flow velocities. This will be useful in \textit{analog gravity} setups which imitate the warp metric \cite{Fischer_2003, finazzi2012analogue} and can be studied experimentally \cite{PhysRevLett.105.240401}. Throughout the paper we have used natural units $G=c=1$ and metric signature $(-,+,+, \cdots)$. 

The rest of the paper is organised as follows. In section \ref{sec:MPchart} we review and extend the construction of MP--charts for spacetimes of the form \eqref{eq: metric} in $D$ dimensions. Interpretation of the \textit{flow coordinates} are discussed including the special case of PG--coordinates. Section \ref{sec:2} highlights the conical singularities of spatial hypersurfaces in three dimensional Minkowski and AdS spacetimes and their conical defect cousins. Regularisation of the singularities and embedding of the warp drive in these spacetimes are discussed. Various properties, like tilting of light cones, event horizons and the violation of (NEC) along with total ($-ve$) energy estimates are explored. We conclude in section \ref{sec: discuss} with comments on possible explorations to higher dimensions, analogue gravity, quantum effects and optimisations. The paper ends with the relevant references. 

\section{Martel--Poisson charts}
\label{sec:MPchart}
For a large class of static spherically symmetric spacetimes in $D\geq 3$ dimensions the metric takes the form 
\begin{equation}
\label{eq: metric}
ds^2=-f d t^2+\frac{d r^2}{f}+r^2 d \Omega_{(D-2)}^2
\end{equation}
where $d \Omega_{(D-2)}^2$ is the line element on an unit $(D-2)$ sphere and $f(r)=(1+\tilde{f}(r))$ \footnote{See later sections \ref{sec:2} and \ref{sec: discuss} for generalisations.}. The main idea is to describe the spacetime in terms of radially infalling or outgoing timelike geodesic observers with D--velocities $ u^\mu=\frac{dx^\mu}{d\tau}=(\dot{t},\dot{r}, 0, \cdots, 0)$ where $\tau$ is the proper time along the geodesic. For static spacetimes, the Killing vector $\xi=(\partial/\partial t)$ leads to a conserved energy per unit mass $\epsilon =-\xi_\mu u^\mu = f \dot{t}$ and utilising the normalisation of the D--velocities $u_\mu u^\mu = -f \dot{t}^2 + (\dot{r}^2/f) = -1 $, we can finally write 
\begin{equation}
\label{eq: co_vel}
u_\mu =\left(-\epsilon, \mp \frac{\sqrt{\epsilon^2-f}}{f}, 0, \cdots, 0 \right) \, , 
\end{equation}
where the ($-ve$) and ($+ve$) signs are for the infalling and outgoing geodesics respectively. 

So far, we have not established that these D--velocities are indeed geodesics. We demand that this congruence of D--velocities are \textit{hypersurface orthogonal} to spacelike surfaces $T(t, r)$ = constant i.e. $u_\mu =-(\partial T/\partial x^\mu)$ where $T$ is now the proper time for these free falling (geodesic) observers. Redefining the time coordinate as \footnote{Note that $T$ is an exact differential i.e. $\displaystyle \frac{\partial \epsilon}{\partial r}=\frac{\partial \left(\frac{\sqrt{\epsilon^2 -f}}{f}\right)}{\partial t}\,$.}
\begin{equation}
\label{eq: PGtime}
dT = \epsilon \vt dt \pm \frac{\sqrt{\epsilon^2 -f}}{f} dr \, ,
\end{equation}
the metric in \eqref{eq: metric} takes the form 
\begin{equation}
\label{eq:MPmetric}   
ds^2 =-dT^2 +  \frac{1}{\epsilon^2}\left(dr + v(r)\vt dT \right)^2 +r^2  d\Omega_{(D-2)}^2 \, ,
\end{equation}
where $v(r)=\pm \sqrt{\epsilon^2-f}$ is the flow velocity. The coordinate charts are valid and cover regions for which $v\in \mathbb{R}$. The geodesic D--velocities of these \textit{Eulerian observers} now take the form 
\begin{equation}
\label{eq: eulerian}
u^\mu =\left(\dot{T}, \dot{r}, 0, \cdots, 0\right) = (1, -v, 0, \cdots, 0) \, .
\end{equation}
Note that except for the case $\epsilon =1$, the induced metric on the hypersurface is not \textit{Ricci flat}. In fact, as we shall see later the scalar curvature and other curvature invariants of the induced metric blows up at $r=0$. 

At this point we can have two different interpretations for the parameter $\epsilon$ which we will now illustrate using the familiar example of Schwarzschild black holes in $D$ dimensions where $\displaystyle f(r)=\left(1-\frac{r_h}{r^{(D-3)}}\right)$, $r_h$ being the location of the horizon and the Eulerian observers are infalling.
\subsubsection*{\textbf{Free fall from finite radius}} 
From the D--velocity we have a conservation equation 
\begin{equation}
\label{eq: energy}
\frac{1}{2} \left(\frac{dr}{dT}\right)^2-\frac{1}{2}\left( \frac{r_h}{r^{(D-3)}} \right) =\frac{\epsilon^2-1}{2}= E \, ,
\end{equation}
where $E$ is the total K.E. + P.E. of the observer. We can choose the observers to start at a finite radius $r_0$ with zero velocity, then $\displaystyle E=-\frac{1}{2}\frac{r_h}{ r_0^{(D-3)}}$ but the coordinate chart is not valid beyond $r>r_0$ as $v(r)$ becomes imaginary. However, if the observers start at infinity then $E=0$ or $ \epsilon=1 $ which gives the PG--coordinates for the Schwarzschild black holes where the induced metric on $T$ = constant hypersurfaces are flat.

\subsubsection*{\textbf{Free fall with initial velocity}}
The observers can start at infinity but now with some initial velocity $v_\infty$ w.r.t. the time coordinate  $t$ such that 
$\displaystyle \epsilon=1/\sqrt{1-v_{\infty}^2} \geq 1$ . Again, zero initial velocity gives the PG--coordinates for the Schwarzschild black holes. There is another interesting limit where we can set $v_\infty=1$ i.e. the speed of light. Scaling $\tilde{T} = (T/\epsilon)$ in \eqref{eq: PGtime} and then taking the limit $\epsilon \rightarrow \infty$ gives the infalling Eddington--Finkelstein (EF) coordinates \cite{Faraoni2020WhenPC}. Of course, it is also possible to have finite radius non--zero initial velocity observers but for the remainder of this paper we shall focus on the initial velocity interpretation for the parameter $\epsilon$.

\section{Warp Drive in various spacetimes}
\label{sec:2}
While the original Alcubierre warp drive is a deformation of the Minkowski spacetime, we shall now discuss embedding the drive in various spherically symmetric static spacetimes, see also \cite{garattini2023black, garattini2024wormholewarp}. Our focus will be on the minimal non--trivial dimension $D=3$ from which it is relatively clear how to generalise for higher dimensions. Three dimensions is also special as it admits \textit{Conical defect} solutions which change the global properties of the spacetime while retaining the local ones. 

\subsection{\textbf{Minkowski spacetime}}
\label{sub: mink}
The MP--charts for the Minkowski spacetime in three dimensions with a tuneable parameter $\epsilon$ takes the form,
\begin{equation}
\label{eq: mink}
ds^2 =-dT^2 +  \frac{1}{\epsilon^2}\left(dr + \sqrt{\epsilon^2-1}\, dT \right)^2 +r^2 d\phi^2 \, , 
\end{equation}
where the radially infalling geodesic observers have the velocity $\displaystyle v_\infty=\sqrt{1-1/\epsilon^2}$ at spatial infinity.
Though this metric has non--vanishing Christoffel symbols, it is just an artefact of the coordinate choice and the curvature tensors vanish. 
\subsubsection*{\textbf{Conical singularity}}
It is instructive to analyse the induced metric on the $T$ = constant hypersurfaces,
\begin{equation}
\label{eq: cone}
ds^2_{C_\epsilon} =\frac{1}{\epsilon^2}dr^2 +r^2 d\phi^2 \quad 0\leq \phi < 2 \pi \, .
\end{equation}
By carrying out the substitutions $r= \epsilon \tilde{r}$ and $\phi = \epsilon \tilde{\phi}$ such that $0\leq \tilde{\phi} < 2 \pi \epsilon\,$ it is straightforward to realise that this is a metric on a conical space $C_\epsilon$ with a \textit{deficit angle} $2 \pi(1-\epsilon)$. Like the plane, the cone is  everywhere flat except at the tip where its curvature $R$ is singular. Calculations by means of the standard formulas
of the Riemannian geometry based on the existence of \textit{tangent spaces} would fail to reveal this delta--like singularity. A natural recipe to handle such singularities and their generalisations can be found in \cite{Fursaev_1995}. To summarise, we consider an embedding of $\mathcal{C}_\epsilon$ in a three dimensional Euclidean space with the parametrisation $x=\epsilon \vt \tilde{r} \cos (\tilde{\phi} / \epsilon), \, y=\epsilon \vt \tilde{r} \sin (\tilde{\phi} / \epsilon),\, z=\sqrt{\left|1-\epsilon^2\right|} \vt \tilde{r}$ defining the conical surface
\begin{equation}
\label{eq: embcone}
z^2-\frac{\left|1-\epsilon^2\right|}{\epsilon^2}\left(x^2+y^2\right)=0 \quad, \quad z \geq 0 \quad .
\end{equation}
For $\epsilon \neq 1$, there is a singularity at $z=0$ where we can't introduce a tangent space and calculate the curvature in the usual way.
 
We can however, introduce a regularisation by \textit{rolling off} the cone tip. The simplest one corresponds to changing $\mathcal{C}_\epsilon$ to a \textit{hyperbolic space} with a tuneable regularisation parameter $a$,
\begin{gather}
z^2-\frac{\left|1-\epsilon^2\right|}{\epsilon^2}\left(x^2+y^2\right)=a^2 \quad, \quad z \geq 0, \\
d s_H^2=\frac{\tilde{r}^2+a^2 \epsilon^2}{\tilde{r}^2+a^2} d \tilde{r}^2+\tilde{r}^2 d \tilde{\phi}^2 \, \label{eq: hbolic} .
\end{gather}
To evaluate the curvature scalar, we first evaluate the integral curvature on the regularised hyperbolic space in the $a\rightarrow 0$ limit,
\begin{equation}
\label{eq: int_R}
\lim _{H \rightarrow C_\epsilon} \int_{H} R=4 \pi(1-\epsilon)+\int_{C_\epsilon -\{\tilde{r}=0\}} R \,  ,
\end{equation}
where the second integral is zero. In a more general setting one can embed the conical defect in a non Ricci flat two dimensional manifold and the second integral would collect the curvature around the defect. We note that since only the singular point $\tilde{r}=0$ can give rise to the first term in \eqref{eq: int_R}, it is best to introduce a local representation for the curvature on $C_\epsilon\,$,
\begin{equation}
\label{eq: local_R}
{ }^{(\epsilon)} R=\frac{2(1-\epsilon)}{\epsilon} \delta(\tilde{r})+R \, .
\end{equation}
\subsubsection*{\textbf{The warp drive}}
We can embed the Alcubierre warp drive in the infalling Minkowski spacetime background \eqref{eq: mink}. In \cite{Bobrick_2021}, various warp drives discussed in the literature are shown to be related to each other by coordinate transformations. It is crucial to note that the embedding discussed in this paper is not a coordinate transformations of the warp drive in the usual $\epsilon =1$ charts. Therefore, it represents an infinite class of different embeddings parameterised by $\epsilon$ or equivalently $v_\infty$. 

We begin, by writing the metric in the standard ADM form in the Cartesian coordinates (warp drive breaks the spherical symmetry),
\begin{equation}
\label{eq: ADM}
d s^2=-N d T^2+ \gamma_{ij}\left(d x^i +N^i d T\right)\left(d x^j +N^j d T\right) \, ,
\end{equation}
where $N$ is the \textit{lapse}, $N^i$ are the \textit{shift vectors} and $\gamma_{ij}$ are the components of the induced metric on the $T$ = constant spacelike hypersurfaces. For the metric as in \eqref{eq: mink}, the embedding is as follows,
\begin{align}
N &=1 \label{eq: lapse1}\\
N^i &= (1-f(r_s))\sqrt{(\epsilon^2-1)}\,\frac{x^i}{r} - \delta^i_x\, f(r_s)\, v \label{eq: lapse2}\\
\gamma_{ij} &= \delta_{ij} +\left(\frac{1}{\epsilon^2}-1\right) \frac{x^i x^j}{r^2} \label{eq: lapse3}\\
\gamma^{ij} &= \delta^{ij} +\left(\epsilon^2-1\right) \frac{x^i x^j}{r^2} \label{eq: lapse4}\, ,
\end{align}
where for three dimensions, $ r=\sqrt{x^2 +y^2}$ is the radial coordinate, $r_s=\sqrt{(x-x_0(T))^2+y^2}$ is the distance from the centre of the drive which is moving along the ($+ve$) $x$--direction with constant velocity $v$. Here, the \textit{form function} $f(r_s)$ decides the size and shape of the warp drive bubble, its exact form is of little consequence (unless one is optimising) as long as it has the value $f(0)=1$ at the centre of the bubble, $f \rightarrow 0$ as $r \rightarrow \infty$ outside the bubble and drops sharply from $1$ to $0$ at the wall of the bubble at some distance $R$ from the centre of the bubble \footnote{\label{fn:form}One such specific function popular in the literature is 
\begin{equation}
\label{eq: form}
f(r_s)=\frac{\tanh [\sigma(r_s+R)]-\tanh [\sigma(r_s-R)]}{2 \tanh (\sigma R)} \, ,
\end{equation}
where $\sigma$ has the interpretation of being inversely proportional to the bubble wall thickness and can be taken to be large.}. The choice of shift vectors $N^i$ are inspired by \cite{ellis2004warp}, where the flow velocity is that of the warp drive inside the bubble and that of the background spacetime outside the bubble. The Eulerian observers with geodesics $ u_\mu =(-1,0,0)$ are still in free fall and observe no time dilation. Hence, the above metric can be considered as a warp drive different from the \textit{Nat\'{a}rio type} \cite{Nat_rio_2002}. We assume the warp drive to be a test particle moving along the timelike curve $x=x_0(T)=vT$, regardless of the value of $v(T)$. One can verify that the proper time along this curve equals the coordinate time. Note that here, the induced metric is not flat and as discussed earlier will exhibit conical singularity at the origin. Warp drives with non flat $\gamma_{ij}$ have been discussed earlier in \cite{Broeck_1999, garattini2024wormholewarp}. As expected, for $\epsilon=1$, we get back the original Alcubierre warp drive \cite{Alcubierre_1994}. 

For the metric \eqref{eq: mink} which is just the $\epsilon \geq 1$, MP--chart for the Minkowski spacetime, we see that the light cones in the $r$--direction are $dr/dT=\pm 1 -v_\infty$ i.e. they have a tilt proportional to $v_\infty$. This carries over to the generic form of the warp drives \eqref{eq: ADM}, only now the light cone along the $x$--direction has the tilt which can be a complicated function of the spatial metric and the lapse vectors, but at large $r$ takes a simplified form, 
\begin{equation}\label{eq: tilt}
\frac{dx}{dT}=\pm 1 -(1-f(r_s)) v_\infty +v \sqrt{(1-v^2_\infty)} f(r_s) \,.  	
\end{equation}
The warp drives travels along a timelike path inside this locally tilted light cones without violating anything in relativity, as the bubble is just a specific choice of the curvature of spacetime itself and not  a massive object obeying the geodesic equation. Globally, the path taken by the warp drives can be spacelike for high enough warp velocity $v$.

We can also analyse the event horizons for these warp drives in a similar fashion. In order to modify the geometry of the warp bubble from inside i.e. to accelerate or to stop the spaceship, the crew must be able to causally influence the bubble wall in front of them by sending say a photon forward towards the bubble wall. In the reference frame of an observer on the spaceship, $x'\equiv x-x_0(T)$, the forward photon has the trajectory,
\begin{equation}\label{eq: horizon}
\frac{dx'}{dT}=1 -(1-f(r_s)) \left(v_\infty +v \sqrt{(1-v^2_\infty)}\right) \,. \end{equation}
At the centre of the bubble $f(r_s)=1$ and photon travels with the speed of  light but if $\left(v_\infty +v \sqrt{(1-v^2_\infty)}\right)>1$, then at some point inside the bubble $x=x_h$ where $f(x_h)=1-\frac{1}{\left(v_\infty +v \sqrt{(1-v^2_\infty)}\right)}$, the photon stops and remain forever at that point. Thus, there is a horizon at $x=x_h$ such that the outer edge of the bubble is outside the causal future $\mathcal{J}^{+}$ of the spaceship. A warp drive beyond a certain velocity can't be controlled, though in principle it is possible that the entire trajectory of the bubble may be constructed beforehand. It is relatively straightforward to extend these arguments to the conical defect backgrounds and the warp drives in AdS/dS spacetimes discussed in the later sections. 

Few comments on the distance $r_s$ and the induced metric $\gamma_{ij}$ are in order. We can always choose $r_s$ to be the coordinate distance $r_s=\sqrt{(x-x_0(T))^2+y^2}$ from the center of the warp drive. In general, the induced metric $\gamma_{ij}$ is not $\delta_{ij}$, therefore the size \& shape of the bubble inside the boundary $r_s=R$ will evolve over time. Even without the warp drive, the metric in \eqref{eq: mink} has non--zero expansion/contraction and shear for a spacelike patch orthogonal to the congruence of Eulerian geodesic observers \footnote{Non zero \textit{rotation} is not possible as Eulerian geodesics are hypersurface orthogonal.}. Our focus will be on the large $x_0$ and $(\epsilon-1)$ small limit where these background expansion and shear are at check. However, there are other choices, either $r_s=\sqrt{\gamma_{ij}(x-x_0(T))^i (x-x_0(T))^j}$ or $r_s=\text{Geodesic}(x,x_0(T))$ where $\text{Geodesic}(x,x_0(T))$ is the geodesic distance between the center of the warp drive $(x_0(T),0)$ and some point $(x,y)$. Another possible choice for the induced metric is $\gamma_{ij} = \delta_{ij} +(1-f(r_s))\left(\frac{1}{\epsilon^2}-1\right) \frac{x^i x^j}{r^2} \, $, an interpolating metric which is $\delta_{ij}$ inside the bubble and $\gamma_{ij}$ as  in \eqref{eq: lapse3} outside \footnote{It is difficult to implement the other choices of $r_s$ with this choice of induced metric as now the definition of $r_s$ depends on $r_s$ itself through the function $f(r_s)$. This puts severe constraints on possible $f(r_s)$, which we checked to be incompatible with the desired properties of $f(r_s)$.}. We will discuss some of these cases in \ref{app: r_s}. 

With our first choice of $r_s$ and the induced metric \eqref{eq: lapse3} we shall now discuss the movement of the warp bubble due to the expansion and contraction of the space around it. The space expands behind the warp bubble and contracts in front of it, thereby pushing the bubble forward (and maybe a spaceship inside it) at arbitrary velocity $v$ \footnote{The 
 same mechanism also applies to the the expanding universe, wherein
galaxies move faster than light w.r.t. each other due to the expansion of the space itself.}. We shall also discuss the energy density as measured by the Eulerian observers i.e. $\rho =T_{\mu \nu}u^\mu u^\nu$ where $T^{\mu \nu}$ is the \textit{stress--energy tensor} of the metric and $u^\mu=(1, -N^x,-N^y)$ are the 3--velocities.

To evaluate the expansion $\theta$ and energy density $\rho$ associated with the warp drive spacetime, we shall compute at first the \textit{extrinsic curvature} $K_{\mu \nu}= \nabla_\mu u_\nu = \frac{1}{2}\mathcal{L}_u P_{\mu \nu}$ for the $T$ = constant hypersurfaces embedded in the full spacetime. Here, the lapse is $N=1$, the timelike geodesics $u^\mu$ are hypersurface orthogonal and $P_{\mu \nu}=g_{\mu \nu} +u_\mu u_\nu$ is the \textit{projector} to the hypersurfaces. The \textit{pullback} of $K_{\mu \nu}$ to the above hypersurfaces are
\begin{align}
\label{eq: ext}
K_{ij} = \frac{1}{2} \mathcal{L}_u \gamma_{ij} &= \frac{1}{2} \left( \partial_T -\mathcal{L}_N \right)\gamma_{ij} \\
& = \frac{1}{2}\partial_T \gamma_{ij} -\frac{1}{2}\left(\gamma_{kj}D_i N^k + \gamma_{ik} D_j N^k \right) \nonumber \, , 
\end{align}
where $D_i$ are the covariant derivatives w.r.t. the induced metric $\gamma_{ij}$. The expansion is the trace $\theta =K^i_i$. The Einstein tensor $G_{\mu \nu} = R_{\mu \nu} - \frac{1}{2} R\, g_{\mu \nu} $ along with the \textit{Gauss--Codazzi relations} \cite{Gourgoulhon:2012ffd} which connects curvature tensors of the metric to curvatures on the hypersurface(s) give, 
\begin{equation}
\begin{aligned}
\label{eq: gauss}
{ }^{(2)} R+\left(K^2-K^{i j} K_{i j}\right) &= 2\, G_{\mu \nu} u^\mu u^\nu = 16 \pi \rho + 2\Lambda \\
D^k K_{ik}-D_i K &=8 \pi f_i
\end{aligned}
\end{equation}
where $\rho$ is the matter energy density and $f_i$ are the matter momentum density and $\Lambda$ is the cosmological constant. 

For the background metric in \eqref{eq: mink} without the warp drive, the extrinsic curvature and expansion away from the conical singularity at $r=0$ are given by \footnote{The regularisation mentioned in \eqref{eq: hbolic} needs modification as now we want ${ }^{(2)} R+\left(K^2-K^{i j} K_{i j}\right)$ to vanish at $r=0$ as $\rho=\Lambda=0$. The regularised metric around $r=0$ with $0\leq\phi <2 \pi$ is given by
\begin{gather}
\label{eq: reg}
ds^2 =-dT^2 +  \frac{1}{\epsilon^2}\left(A(r) \, dr + \sqrt{\epsilon^2-1}\,B(r)\, dT \right)^2 +r^2 d\phi^2  \nonumber \\
A(r) = \sqrt{\frac{a^2 \epsilon ^4+ r^2}{a^2 \epsilon ^2+ r^2}} \,\,\,\,\, 
B(r) = \frac{\sqrt{\epsilon ^2 (1-2\ln A(r))-1}}{\sqrt{\epsilon ^2-1}}
\end{gather}
}
\begin{equation}
\label{eq: mink_K}
K_{i j}= \sqrt{\epsilon ^2-1} 
\begin{pmatrix}
\frac{-y^2}{r^3} & \frac{x y}{r^3} \\[2pt]
 \frac{x y}{r^3} & \frac{-x^2}{r^3} \\
\end{pmatrix} \, , \quad \theta = -\frac{\sqrt{\epsilon ^2-1}}{r} \, .
\end{equation}
Closed form expressions for the $K_{ij}$ and $\rho$ can be obtained for the metric with the warp drive centred at $x_0$. However, we are interested in the limit $x_0 \gg (x-x_0) \approx y \approx r_s \,$, where the warp drive is sufficiently far way from the centre $r=0$. We shall also take $v_\infty \ll 1$ for the Eulerian observers. To $\mathcal{O}\left(\frac{1}{r}\right)$ and $\mathcal{O}(v_\infty)$, the expressions are as follows \footnote{Here, ${}^{(2)} R =0$ on the hypersurfaces, except at $r=0$. This is not true in $D \geq 4$ dimensions, thereby opening up new possibilities to minimise the energy density.}
\begin{align}
\theta = & \,(v+v_\infty)(x-x_0)\frac{f'(r_s)}{r_s} -(1-f(r_s))\,\frac{v_\infty}{r} \nonumber \\ 
& +\frac{y^2 \,v_\infty}{r_s r} f'(r_s) \\
\rho = & -\frac{1}{16 \pi} \left [\frac{1}{2} (v+v_\infty)^2 \, y^2 \frac{f'(r_s)^2}{r_s^2} \right. \label{eq: wrap_mink} \\
 & \left. +v\, v_\infty (x-x_0) \left(2 \,r_s(1-f(r_s)) - y^2 f'(r_s)\right) \frac{f'(r_s)}{{r_s^2\, r}}\right ]. \nonumber 
\end{align}

At first, we note that for $v_\infty=0$, the expressions falls back to the usual Alcubierre warp drive in the Minkowski spacetime. As expected, the $1/r$ zeroth order expressions for both $\theta$ and $\rho$ have the combination $(v+v_\infty)$ for velocity. In these \textit{river flow coordinates}, the background metric is \textit{flowing} like a fluid along the radially inward direction. However, the warp drive is flowing against this ``fluid'' in the $(+ve)$ $x$--direction, i.e. it is flowing with velocity $(v+v_\infty)$ w.r.t. the fluid. Therefore, it is best to move along the flow to minimise the peak expansions/contractions and $(-ve)$ energy density. In fact, for a free falling warp drive to zeroth order they are zero but picks up a ($-ve$) contribution at $\mathcal{O}\left(\frac{1}{r^2}\right)$. We also note that, the first $\mathcal{O}\left(\frac{1}{r}\right)$ term in $\theta$ is that of the background metric and maybe subtracted, the other one is proportional to $(y/r)$ and hence very small. As $f'(r)<0$, behind the bubble $\theta>0$ and the space is expanding, while at the front $\theta<0$ and the space is contracting. By construction $f'(r_s)$ is non zero only along a very small width of the ring $r_s=R$. Hence, the effects of the warp drive in both $\theta$ and $\rho$ are localised around this ring, the boundary of the warp bubble.
\subsubsection*{\textbf{Conical defect backgrounds}}
Ordinary gravity in three dimensions is special, it is dynamically trivial.  Effects of localised sources are on the global properties whereas the outside metric is flat. For a point source of mass $M$ placed at the origin $r=0$, the spatial metric is a cone and the full metric is given by 
\begin{equation}
\label{eq: conical_mink}
ds^2 = -dt^2 +\frac{1}{\alpha^2} \vt dr^2 + r^2 d\phi^2 \quad 0\leq\phi<2 \pi \, ,
\end{equation}
where $\alpha = 1-4M$ \cite{DESER1984220} \footnote{Owing to \textit{Birkhoff's theorem}, in $D\geq4$ dimensions, point mass solutions with spherical symmetry are Schwarzschild like. We can however break the spherical symmetry in solutions like the cosmic string or p--branes where the mass density is concentrated along a line or a membrane and the traverse spatial metric is still conical \cite{Kibble:1976sj, Randall_1999}.}. We would like $\alpha$ to be positive which restricts the mass $M<(1/4)$. Negative mass or $\alpha>1$ can be considered as  unphysical. It is instructive to scale $t\rightarrow \alpha \vt t$ and write the metric as MP--charts with the parameter $\epsilon$,
\begin{equation}
\label{eq: conical_mink_poi}
ds^2 = -dT^2 + \frac{1}{\epsilon^2} \left(dr +\sqrt{\left(\epsilon^2-\alpha^2\right)}\, dT\right)^2 + r^2 d\phi^2\, , 
\end{equation}
such that $v_\infty = \alpha^2\sqrt{\left(1-\frac{1}{\epsilon^2}\right)\,}$. Keeping in mind that the charts are invalid for $\left(\epsilon^2-\alpha^2\right)<0$, there are two natural choices for the parameter $\epsilon$ to realise an embedding of the warp drive in these defect backgrounds. 

\paragraph{\textbf{Case $\epsilon =1$}}: Though $v_\infty$ is zero, the flow velocity of the background metric is non--zero. However, the spatial metric $\gamma_{ij}=\delta_{ij}$ and non--singular at the origin $r=0$. The expressions for the $\theta$ and $\rho$ are as follows,
\small
\begin{align}
\label{eq: wrap_cone_mink_1}
\theta = & \left(v+\sqrt{1-\alpha^2}\right)(x-x_0)\frac{f'(r_s)}{r_s}  \\
  & -(1-f(r_s)) \frac{ \sqrt{1-\alpha^2}}{r} +\sqrt{1-\alpha^2}\,y^2\frac{f'(r_s)}{r_s r} +\mathcal{O}\left(\frac{1}{r^2}\right) \nonumber \\
\rho = & -\frac{1}{16 \pi} \left [\frac{1}{2} \left(v+\sqrt{1-\alpha^2}\right)^2\frac{f'(r_s)^2}{r_s^2} + \mathcal{O}\left(\frac{1}{r}\right) \right] \, .
\end{align}
\normalsize
Here, the first $\mathcal{O}\left(\frac{1}{r}\right)$ term in $\theta$ is the background metric contribution and may be subtracted. Again the $(-ve)$ energy density is localised around the boundary of the warp bubble, even for the $\mathcal{O}\left(\frac{1}{r}\right)$ terms which we omit writing down. We also note that in this conical defect background the the light cone tilt as given in \eqref{eq: tilt} is modified to
\begin{equation}
\frac{dx}{dT}=\pm 1 -(1-f(r_s)) \sqrt{1-\alpha^2} +v f(r_s) \,, 	
\end{equation}
and the forward horizon \eqref{eq: horizon}, is now controlled by the equation
\begin{equation}
\frac{dx'}{dT}=1 -(1-f(r_s)) \left(\sqrt{1-\alpha^2} +v \right) \,. 	
\end{equation}
At some point inside the bubble $x=x_h$ where $f(x_h)=1-\frac{1}{\left(\sqrt{1-\alpha^2} +v \right)}$, a horizon develops even for $1-\sqrt{1-\alpha^2}\leq v<1$ . This should be contrasted with the original Alcubierre warp drive in Minkowski space where the horizon forms only when $v\geq 1$.
\paragraph{\textbf{Case $\epsilon =\alpha$}}:  Here, $v_\infty= \alpha \sqrt{\left(\alpha^2-1\right)}$. This is only possible if we allow the defect to be sourced by a $(-ve)$ mass $M$. This is a particularly simple case where the flow velocity of the background metric vanishes. The expressions for the $\theta$ and $\rho$ are as follows,
\begin{align}
\label{eq: wrap_cone_mink_2}
\theta = & v\vt (x-x_0)\frac{f'(r_s)}{r_s}\\
\rho= & -\frac{1}{16 \pi} \left [\frac{1}{2} v^2y^2 \frac{f'(r_s)^2}{\alpha^2\, r_s^2} \right. \\
& \left. + \vt v^2(x-x_0)y^2(\alpha^2-1)\frac{f'(r_s)^2}{\alpha^2\vt r_s^2\vt r} + \mathcal{O}\left(\frac{1}{r^2}\right) \right ] \, .\nonumber 
\end{align}
We note that the expansion $\theta$ is the same as the original Alcubierre warp drive whereas the $(-ve)$ energy is suppressed by a factor of $(1/\alpha^2)$ \footnote{ In this case the light cone tilt is given by $\frac{dx}{dT}=\pm 1 +\frac{v}{\alpha} f(r_s)$ and the forward horizon is at $f(x_h)=1-\frac{\alpha}{v}$ i.e. a horizon forms for all $v\geq\alpha$ . }.

\subsubsection*{\textbf{Energy considerations}}
So far, we have shown that the warp drives in the three dimensional Minkowski spacetime and its conical defect cousins in $\epsilon \geq 1$, MP--charts carry ($-ve$) energy densities localised around the warp bubble which means they violate the \textit{weak energy condition} (WEC). Matt Visser \textit{et. al.} in \cite{Santiago_2022} argued that for almost all Nat\'{a}rio type warp drives discussed in the literature with a generic line element of the form,
\small
\begin{equation}
ds^2=-dt^2+\delta_{i j}\left(dx^i-v^i(\vec{x}, t) dt\right)\left(dx^j-v^j(\vec{x}, t)  dt\right) \, ,	
\end{equation}
\normalsize
including the ones that claim ($+ve$) energy densities observed by the Eulerian observers violate either the \textit{strong energy condition} (SEC), \textit{dominant energy condition} (DEC), (WEC) or the \textit{null energy condition} (NEC) for a generic observer. However, their analysis  does not cover the warp drives discussed in this paper as the spatial metric $\gamma_{i j} \neq \delta_{i j}$. In general, the violation of the energy conditions follow the linkage \cite{Kontou_2020},
\begin{equation}
\mathrm{NEC}\longrightarrow \mathrm{WEC} \longrightarrow \mathrm{DEC} \quad \text{and} \quad \mathrm{NEC} \longrightarrow \mathrm{SEC}\, .
\end{equation}
 As we shall see, for the warp drives discussed here, (NEC) is violated which leads to the violation of the other energy conditions.
 
 In order to explore the violation to the (NEC) it is best work with the natural co--moving orthonormal frame (triad) attached to the Eulerian observers of the the warp drive metric \eqref{eq: ADM} with $N=1$. For the spatial metric $\gamma_{ij}$, we can choose the spatial dyad $\tilde{e}^{\hat{j}}_i$ such that $\displaystyle \gamma_{ij}=(\tilde{e}^{\hat{i}})_i (\tilde{e}^{\hat{j}})_j \delta_{\hat{i} \hat{j}}$ . Then the triads for the metric $g_{\mu \nu}$ are 
 \begin{equation}
  \begin{aligned}
 	(e^{\hat{0}})_\mu &= u_\mu=(-1;0,0)_\mu \\
 	 (e^{\hat i})_\mu &= (N^j (\tilde{e}^{\hat{i}})_j;(\tilde{e}^{\hat{i}})_1,(\tilde{e}^{\hat{i}})_2)_\mu \\
 	(e^{\hat{0}})^\mu &=u^\mu =(1;-N^1,-N^2)^\mu \\(e^{\hat{i}})^\mu &=(0;\gamma^{1 j}(\tilde{e}^{\hat{i}})_j,\gamma^{2 j}(\tilde{e}^{\hat{i}})_j)^\mu \, .
 \end{aligned}
 \end{equation}
 Usually it is assumed that the stress--energy tensor is of the \textit{Hawking--Ellis type I} \cite{Hawking_Ellis_1973} which means it is diagonal in an orthonormal basis for the observer. Unfortunately, for the warp drive solutions we do not know a priori whether or not the stress--energy is Hawking--Ellis type I in general, nor do the Eulerian observers diagonalise the stress--energy tensor. This requires modification to the usual approach to testing energy conditions.
 
 The null energy condition (NEC) demands that for all null vectors $k^\mu$, we must have $T_{\mu \nu} k^\mu k^\nu \geq 0$ . In the orthonormal frame, let us take any two oppositely oriented null vectors $k_{+}^{\hat{\mu}}=(1,+k^{\hat i})$, and $k_{-}^{\hat{\mu}}=(1,-k^{\hat i})$ where $k^{\hat i}$ is an arbitrary unit spatial 2--vector. Then the (NEC) would demand,
 \begin{equation}
 \begin{aligned}
 T_{\hat{\mu} \hat{\nu}} k_{+}^{\hat{\mu}} k_{+}^{\hat{\nu}}&=\rho+2 T_{\hat{0} \hat{i}} k^{\hat i}+T_{\hat{i} \hat{j}} k^{\hat i} k^{\hat j} \geq 0 \\
 T_{\hat{\mu} \hat{\nu}} k_{-}^{\hat{\mu}} k_{-}^{\hat{\nu}}&=\rho - 2 T_{\hat{0} \hat{i}} k^{\hat i}+T_{\hat{i} \hat{j}} k^{\hat i} k^{\hat j} \geq 0 \, ,
\end{aligned}
\end{equation}
which upon averaging implies the following one--way relation, $\mathrm{(NEC)} \longrightarrow \rho+T_{\hat{i} \hat{j}} k^{\hat i} k^{\hat j} \geq 0$ .
 For a dyad $k_A^{\hat i}$ of two mutually orthogonal unit vectors, for each $A$ (NEC) implies, $
 \rho+T_{\hat{i} \hat{j}} k_A^{\hat i} k_A^{\hat j} \geq 0 $ . Noting that by construction $\sum_A k_A^{\hat i} k_A^{\hat j}=\delta^{\hat{i} \hat{j}}$ and averaging over the two members of the dyad gives 
 \begin{equation}
\rho+T_{\hat{i} \hat{j}}\left(\frac{1}{2} \sum_A k_A^{\hat i} k_A^{\hat j}\right) \geq 0 \quad \text{or} \quad \rho+\frac{1}{2} T_{\hat{i} \hat{j}}\delta^{\hat{i} \hat{j}} \geq 0 \, .
\end{equation}
Even if the 2--stress tensor $T_{\hat{i} \hat{j}}$ is not diagonal, we can define the average pressure as usual, $\bar{p}=\frac{1}{2} T_{\hat{i} \hat{j}} \delta^{\hat{i} \hat{j}}$ . Finally, we have the one--way relation,
\begin{equation}
\mathrm{(\textbf{NEC})} \longrightarrow \rho+\bar{p} \geq 0 \, .	
\end{equation}
Similarly, for the other energy conditions, the implications are \footnote{There are other energy conditions like \textit{trace energy condition} (TEC), \textit{flux energy condition} (FEC) etc. and various weaker average energy conditions of which \textit{averaged null energy condition}  (ANEC) is widely applicable.}
\begin{itemize}
	\item[$\bullet$] (\textbf{WEC}) : $\rho\geq 0$ and $\rho+\bar{p} \geq 0$ coming from the demand $T_{\mu \nu} \ell^\mu \ell^{\nu} \geq 0$ for all timelike vectors $\ell^\mu$.
	\item[$\bullet$] (\textbf{SEC}) : $\rho+2\bar{p}\geq 0$ and $\rho+\bar{p} \geq 0$ coming from the demand $\left(T_{\mu \nu}-\frac{1}{2} T g_{\mu \nu}\right) \ell^\mu \ell^\nu \geq 0$ for all timelike vectors $\ell^\mu$.
	\item[$\bullet$] (\textbf{DEC}) : $|\bar{p}|\leq \rho$ coming from the demand $T_{\mu \nu} \ell^\mu \tilde{\ell}^{\nu} \geq 0$ for all future--pointing timelike vectors $\ell^\mu$ and $\tilde{\ell}^{\nu}$.
\end{itemize}

To compute the average pressure $\bar{p}=\frac{1}{2} T_{\hat{i} \hat{j}} \delta^{\hat{i} \hat{j}}=\frac{1}{2} T_{ij} \gamma^{ij}=\frac{1}{2}T_{\mu \nu}P^{\mu \nu}$ for the warp drives discussed in this paper we can use \eqref{eq: gauss} along with other Gauss--Codazzi relations,
\begin{equation}
\begin{aligned}
	R_{ij} &={}^{(2)}R_{ij} +\mathcal{L}_u K_{ij} -2(K^2)_{ij}+K K_{ij} \\
	R_{uu} &=-\mathcal{L}_u K -\mathrm{tr}(K^2) \\
	R_{u i} &=D^k K_{ik}-D_i K \, ,
\end{aligned}
\end{equation}
to evaluate the Einstein tensor $G_{ij}$ and hence $T_{ij}$ . For the warp drive described by \eqref{eq: ADM} and \eqref{eq: lapse1} to \eqref{eq: lapse4}, the energy density is already evaluated in \eqref{eq: wrap_mink}. To check for the violation of (NEC) to the leading order in $r$, we note that the form function $f(r_s)$ is a monotonically decreasing function such that $f'(r_s)<0$, $f''(r_s)<0$ and $0\leq f(r_s) \leq 1$ . Therefore, (NEC) is violated for the $\epsilon\geq1$, MP--charts warp drives,
\small
\begin{align}
\rho +\bar{p}=& -\frac{1}{32 \pi \epsilon^2 r_s^3}\Bigl[\left(v+\sqrt{\epsilon^2-1}\right)^2 \left\{y^2 \epsilon^2(f(r_s)-1)f'(r_s) \right.  \nonumber \\
 & \left. +r_s \left(y^2+\epsilon^2(x-x_0)^2\right)f'(r_s)^2 \right.  \\
 & \left. +r_s \epsilon^2(x-x_0)^2(f(r_s)-1)f''(r_s) \right\}\Bigr] \leq 0 \, . \nonumber 
\end{align}
\normalsize
We can do similar exercises for the warp drives embedded in the conical defect backgrounds, again (NEC) is violated as expected. The (NEC) violation for these warp drive solutions leads to violations of other energy conditions, (WEC), (DEC) and (SEC).

We now turn our focus to give a reasonable estimate of the total energy measured by the Eulerian observers. For the $\epsilon\geq 1$, MP--charts warp drives, shifting the origin to the location of the warp drive and to the leading  order in $r$, the total energy is given by integrating the energy density \eqref{eq: wrap_mink} over a spatial slice,
\begin{equation}\label{eq: totale}
\begin{aligned}
E &=\int T_{\mu \nu}u^\mu u^\nu d^2 x \\
&=-\frac{\left(v+\sqrt{\epsilon^2-1}\right)^2}{32 \pi \epsilon^2} \int \frac{(r \sin{\phi})^2 f'(r)^2}{r^2} r dr d\phi	\\
&= -\frac{\left(v+\sqrt{\epsilon^2-1}\right)^2}{64 \pi \epsilon^2} \int f'(r)^2 r dr \, .
\end{aligned}
\end{equation}
Of course, $E$ is ($-ve$) and depends on the choice of the form function $f(r)$. Instead of choosing the usual $f(r)$ \eqref{eq: form}, we shall use the piecewise--continuous form function suggested by Pfenning and Ford \cite{Pfenning_1997} as it provides better estimates for the total energy,
\begin{equation}
f(r)= \begin{cases}1 & r \leq R-\frac{\delta}{2} \\ \frac{1}{2}+\frac{R-r}{\delta} & r \in\left(R-\frac{\delta}{2}, R+\frac{\delta}{2}\right) \\ 0 & r \geq R+\frac{\delta}{2}\end{cases}
\end{equation}
Here, $R$ is the radius of the warp drive bubble and $\delta$ is its wall thickness. Plugging this into \eqref{eq: totale}, we get
\begin{equation}\label{eq: tote}
\begin{aligned}
E&=	-\frac{\left(v+\sqrt{\epsilon^2-1}\right)^2}{64 \pi \epsilon^2}\int_{R-\frac{\delta}{2}}^{R-\frac{\delta}{2}}\left(\frac{1}{\delta}\right)^2 r dr \\
&=-\frac{\left(v+\sqrt{\epsilon^2-1}\right)^2}{64 \pi \epsilon^2} \frac{R}{\delta} \, .
\end{aligned}
\end{equation}
We can choose a reasonable size for the bubble, say $R=100 \mathrm{~m}$ and make a choice for the background flow velocity ($v_{\infty} $) but the crucial ingredient is to estimate the maximum thickness of the bubble wall so as to reduce the overall ($-ve$) energy to its minimum. 

The restrictions to the thickness of the bubble can be obtained from \textit{quantum inequalities} (QI) which do allow ($-ve$) energies but places serious limitations on its magnitude and duration \cite{PhysRevD.43.3972}. It would be very difficult to quantise a scalar field on a warp drive background and arrive at the exact (QI), but upto corrections in the inverse powers of the local radius of curvature, the flat spacetime (QI) can be applied to curved spacetimes with the additional restriction that the ($-ve$) energy be sampled on timescales smaller than the minimum local radius of curvature \cite{Pfenning_1997_1}. Most of the literature on (QI) is in four dimensions but the warp drives in this paper are in three dimensions. Redoing the computations, we arrive at the expression,
\begin{equation}
\frac{T_0}{\pi} \int_{-\infty}^{\infty} \frac{\left\langle T_{\mu \nu} u^\mu u^\nu\right\rangle}{T^2+T_0^2} d T \geq-\frac{1}{16 \pi T_0^3} \, ,
\end{equation} 
where $g(T)=\frac{T_0}{\pi(T^2+T_0^2)}$ is a suitable \textit{sampling function}.
For the $\epsilon\geq 1$, MP--charts warp drives in consideration we notice that the largest component of the Riemann tensor is given by
\begin{equation}
\bigl|R_{\hat{T} \hat{y} \hat{T} \hat{y}}\bigr|=\frac{3 \left(v+\sqrt{\epsilon^2-1}\right)^2 y^2}{4 \epsilon^2 r^2}f'(r)^2 \, ,
\end{equation}
which yields the minimum radius of curvature to be 
\begin{equation}
r_{\min } \equiv \frac{1}{\sqrt{\bigl|R_{\hat{T} \hat{y} \hat{T} \hat{y}}\bigr|}} \sim \frac{2 \epsilon \delta}{\sqrt{3} \left(v+\sqrt{\epsilon^2-1}\right)}\, ,
\end{equation}
near the boundary of the bubble. As mentioned earlier, the sampling time must be smaller than this length scale, so we can take
\begin{equation}
T_0=\Delta \frac{2 \epsilon \delta}{\sqrt{3} \left(v+\sqrt{\epsilon^2-1}\right)} \quad 0<\Delta \ll 1 \, ,
\end{equation}
where $\Delta$ is an unspecified parameter which we choose to be, $\Delta=1/10$ . Following the steps in \cite{Pfenning_1997}, (QI) puts a  bound on the thickness of the bubble,
\begin{equation}
\delta \leq \frac{3\sqrt{3}}{4 \epsilon \Delta^3}\left(v+\sqrt{\epsilon^2-1}\right)\sim 10^3 \frac{\left(v+\sqrt{\epsilon^2-1}\right)}{\epsilon}\ell_p \, ,
\end{equation}
where $\ell_p$ is the Planck length \footnote{In three dimensions, $\ell_p=\frac{\hbar G_3}{c^3}$ but if we assume that we reach three dimensions by compactifying one spatial direction of order $\ell_p$ in four dimensions, then $G_3=G_4/\ell_p$ and $\ell_p=1.6 \times 10^{-35} \mathrm{~m}$.}. We can assume the mass of a Milky--way galaxy in three dimensional universe to be two third of the actual Milky--way, $M_{\mathrm{Milky}}=1.3 \times 10^{33} ~m_p$ . Putting it all together in \eqref{eq: tote}, we get
\begin{equation}
E\leq -3 \times M_{\mathrm{Milky}} \frac{\left(v+\sqrt{\epsilon^2-1}\right)}{\epsilon}	\, .
\end{equation}
It is about $10^{20}$ times better than the four dimensional result but still   a stupendously large amount of ($-ve$) energy is required to operate the warp drive in three dimensions.   

The energy considerations for the warp drives embedded in the conical defect backgrounds are similar to the above analysis. If we compare with the original $\epsilon=1$, Alcubierre warp drive in three dimensions, the energy density and total energy of the $\epsilon=1$, conical defect warp drives scales by a factor of  $\left(\frac{\left(v+\sqrt{1-\alpha^2}\right)}{v}\right)\geq 1$, assuming that the warp drive is travelling outwards. This is expected as the conical defect background has a flow inwards towards the origin. Similarly, for the other case of $\epsilon =\alpha$, the suppressing ratio is $\frac{1}{\alpha^2}\leq 1$ i.e. it has the potential to lower the energy requirements, but unfortunately it also requires ($-ve$) mass to create the defect itself. The analysis also goes through for the warp drives embedded in the AdS/dS and their conical defect cousins discussed in the next section in the small Hubble constant, $H \ll 1$ limit \footnote{The finite $H$ computations are challenging. Firstly the metric is complicated in the Cartesian coordinates which is necessary to correctly embed the warp drives and secondly the (QI) for AdS/dS spaces would require quantising scalar fields in curved spacetimes. We shall attempt a more complete analysis for these cases in a future work.}. 

\subsection{\textbf{AdS spacetime}}
\label{sub: ads}
As discussed in \cite{Faraoni2020WhenPC} the MP--charts for \textit{anti--de Sitter space} covers only part of the spacetime. The $D$ dimensional metric in global coordinates take the form 
\begin{equation}
\label{eq: ads}
d s^2=-\left(1+H^2 r^2\right) d t^2+\frac{d r^2}{1+H^2 r^2}+r^2 d \Omega^{2}_{(D-2)}
\end{equation}
where $H=\frac{1}{\ell}$ is the Hubble constant, $\ell$ is the AdS length scale and $\Lambda=-\frac{(D-1)(D-2)}{2 \, \ell^2} $ is the cosmological constant. Following the arguments in section \ref{sec:MPchart} one can construct the MP--charts with parameter $\epsilon$, however now the Eulerian observers are outgoing and travels with initial velocity $v_\infty$ from $r=0$ with proper time $T$ defined as, 
\begin{equation}
\label{eq: PGtime_ads}
dT = \epsilon \vt dt - \frac{\sqrt{\epsilon^2 -\left(1+H^2 r^2\right)}}{\left(1+H^2 r^2\right)} dr \, .
\end{equation}
Our focus is mainly on $\text{AdS}_3$ with the line element,
\begin{gather}
\label{eq:MPads}   
ds^2 =-dT^2 +  \frac{1}{\epsilon^2}\left(dr - \sqrt{\epsilon^2 -\left(1+H^2 r^2\right)} dT \right)^2 +r^2  d\phi^2 \nonumber\\
0 \leq r \leq \frac{\sqrt{\epsilon^2-1}}{H} \equiv r_{+} \, ,
\end{gather}
where due to the $(-ve)$ \textit{Misner--Sharp--Hernandez} mass \cite{Misner1964RELATIVISTICEF,Hernandez:1966zia} of AdS spacetimes the usual $\epsilon=1, \,v_\infty=0$, PG--coordinates doesn't exist \footnote{We note that for \textit{de--sitter} dS spacetimes, the metric is as in \eqref{eq: ads} with $H^2\rightarrow -H^2$ and PG--coordinates exists along with the MP--charts covering the full spacetime.}.

Here too, the $T$ = constant hypersurfaces exhibit conical singularity at $r=0$ as discussed in section \ref{sub: mink}. With the assumption that $H\ll 1$ and $\epsilon$ is some reasonable number greater than 1, we can push the  charts to cover large regions of spacetime except for regions near the AdS boundary $r\rightarrow \infty$. The extrinsic curvature tensor away from the singularity is as follows,
\begin{align}
\label{eq: ads_K}
& K_{ij} = - \\
&\begin{pmatrix}
 \frac{H^2 r^2 \left(x^2+y^2 \epsilon ^2\right)-y^2 \epsilon ^2 \left(\epsilon ^2-1\right)}{r^3 \epsilon ^2 \sqrt{\epsilon^2 -\left(1+H^2 r^2\right)}} & \frac{x y \left(\epsilon
   ^2-1\right) \left(\epsilon ^2-H^2 r^2\right)}{r^3 \epsilon ^2 \sqrt{\epsilon^2 -\left(1+H^2 r^2\right)}} \\[3pt]
 \frac{x y \left(\epsilon ^2-1\right) \left(\epsilon ^2-H^2 r^2\right)}{r^3 \epsilon ^2 \sqrt{\epsilon^2 -\left(1+H^2 r^2\right)}} & \frac{H^2 r^2 \left(x^2 \epsilon ^2+y^2\right)-x^2
   \epsilon ^2 \left(\epsilon ^2-1\right)}{r^3 \epsilon ^2 \sqrt{\epsilon^2 -\left(1+H^2 r^2\right)}} \\
\end{pmatrix} \nonumber \\
& \theta = \frac{\epsilon^2 -\left(1+2 H^2 r^2\right)}{r \sqrt{\epsilon^2 -\left(1+H^2 r^2\right)}} \, .
\end{align}

We shall now embed the warp drive centred at $x_0$ in this background. Closed form expressions for the $K_{ij}$ and $\rho$ can be obtained but as in the Minkowski case, we are mostly interested in the limit $x_0 \gg (x-x_0) \approx y \approx r_s \,$, where the warp drive is sufficiently far way from the singularity at the centre $r=0$. As we have finite range charts, we have to treat the order of limits carefully. The radial distance $r$ in these MP--charts is cutoff at $r_{+}(\epsilon, H)\,$ therefore it is best to scale it as $r= (1-s_r)\vt r_{+}$ such that $0\leq s_r \leq 1$ and then focus on the region near $s_r \approx 0\,$ \footnote{Very near to the MP--charts cutoff, the light cone tilt is given by $\frac{d x}{dT}=\pm 1 +v \sqrt{1-v^2_\infty}$ and the forward photon horizon forms at $f(x_h)=1-\frac{1}{\left(v \sqrt{(1-v^2_\infty)}\right)}$ i.e. $\displaystyle v \gtrapprox 1+\frac{1}{2} v^2_\infty$ .}. To $\mathcal{O}(s_r)$ and $\mathcal{O}(H)$, the expansion and energy density are as follows, 
\begin{align}
\theta =& \left(v-\sqrt{2(\epsilon^2-1)\vt s_r}\right)(x-x_0)\frac{f'(r_s)}{r_s} \label{eq: wrap_ads_1} \\
& -\frac{(1-f(r_s)}{\sqrt{2}}\left(\frac{1}{\sqrt{s_r}}-\frac{11}{4}\sqrt{s_r}\right) H \nonumber \\
& -\sqrt{2} \vt y^2 \sqrt{s_r}  \frac{f'(r_s)}{r_s} \vt H \nonumber \\
\rho =& -\frac{1}{16 \pi} \left [\frac{1}{2} \left(v-\sqrt{2\vt (\epsilon^2-1)s_r}\right)^2 y^2 \frac{f'(r_s)^2}{r_s^2 \epsilon^2} \right.  \label{eq: wrap_ads_2}\\
& +v^2 y^2(x-x_0)\sqrt{(\epsilon^2-1)} \frac{f'(r_s)^2}{r_s^2 \epsilon^2} H \nonumber \\
& \left. +\mathcal{O}(\sqrt{s_r})\vt H +\mathcal{O}(H^2) \vphantom{\frac{1}{2}}\right] \, . \nonumber
\end{align}

We notice the combination, $\left(v-\sqrt{2\vt(\epsilon^2-1)\vt s_r}\right)$ appearing at the zeroth order, however for AdS the background flow velocity and the warp drive velocity are both along $(+ve)$ $x$--direction. The first $\mathcal{O}(H)$ term in \eqref{eq: wrap_ads_1} is the contribution from the background AdS and maybe subtracted and the second term is very small. Similarly, in \eqref{eq: wrap_ads_2} the background contribution from the cosmological constant has been subtracted in the $\mathcal{O}(H^2)$ terms. After the appropriate subtractions remaining terms are again proportional to $f'(r_s)^{\#}$ such that the change in $\theta$ and $\rho$ are localised around the boundary of the warp bubble. Interestingly, the zeroth order term in the $(-ve)$ energy density is suppressed by $(1/\epsilon^2)$ thereby reducing the energy as $\epsilon>1$ \footnote{This is also true for the Minkowski space. It is not visible in \eqref{eq: wrap_mink} as we truncated the expressions to $\mathcal{O}(v_\infty)$.}. 

The analysis for warp drive embedded in dS spacetime is similar. However, the MP--charts now cover the full spacetime. Even crossing the  \textit{apparent horizon} at $r=(1/H)$ is regular. To study the warp drive near this horizon, we can use the scaling $r=((1-s_r)/H)$ and follow the steps as discussed above. the results are similar and as expected, nothing special happens because of the horizon.
\subsubsection*{\textbf{Conical defect backgrounds}}
Similar to the case of conical defect solutions in three dimensional Minkowski space, $\text{AdS}_3$ too have conical defect solutions for a point mass $M$ placed at the origin $r=0$. The spatial metric is a cone and the full metric is given by
\begin{equation}
\label{eq: conical_ads}
ds^2 =-(\alpha^2 +r^2H^2)dt^2 +\frac{dr^2}{(\alpha^2 +r^2H^2)} +r^2 d\phi^2 \, ,
\end{equation}
where $\alpha =1-4M$ and $0\leq\phi<2 \pi$ \cite{DESER1984405} \footnote{ $\text{dS}_3$ is  bit tricky. In all three spacetimes, there is a second source at $r=\infty$. In Minkowski and $\text{AdS}_3$ spacetimes the geodesic distance between the source at $r=0$ and $r=\infty$ is infinite, so the second source can be ignored but for $\text{dS}_3$ it is finite and contributes.}. The MP--charts with $\epsilon=\alpha$ doesn't exist. Therefore, we focus on the $\epsilon=1,\, v_\infty=0$ charts given by
\begin{gather}
\label{eq: conical_ads_poi}
ds^2=-dT^2+\left(dr -\sqrt{(1-\alpha^2)-r^2H^2} \,dT\right)^2 + r^2 d\phi^2 \nonumber \\
0\leq r \leq \sqrt{1-\alpha^2}\,H^{-1} \equiv {}^c r_{+} \,.
\end{gather}
We note that for physical particles, we have $0<\alpha\leq1$ \footnote{In the conical AdS background, very near to the MP--charts cutoff, the light cone tilt is given by $\frac{d x}{dT}=\pm 1 +v $ and the forward photon horizon forms at $f(x_h)=1-\frac{1}{v}$ i.e. $v \geq 1$ . This of course gets corrected at $\mathcal{O}(s_r)$ .}. 

We shall now embed the warp drive in this conical $\text{AdS}_3$ background and do a similar analysis as done earlier. The important changes are that the spatial metric is now $\delta_{ij}$ and scaling of the radial coordinate is $r=(1-s_r)\, {}^cr_{+}$ such that $0\leq s_r\leq 1$ and then we focus on the region $s_r\approx 0$. The expansion $\theta$ is as given in \eqref{eq: wrap_ads_1} with the replacement $(\epsilon^2-1) \rightarrow (1-\alpha^2)$. Parts of the $\mathcal{O}(H)$ not proportional to $f'(r_s)$ are from the background metric and should be subtracted. For the energy density $\rho$, the zeroth order $\mathcal{O}(H^0)$ terms is the same as in \eqref{eq: wrap_ads_2} with the replacement $(\epsilon^2-1) \rightarrow (1-\alpha^2)$ while omitting the $\epsilon^2$ in the denominator. From $\mathcal{O}(H)$, terms are different, in this case starting at $\mathcal{O}\left(\sqrt{s_r}\right)$. Again the background contribution from the cosmological constant needs to be subtracted from the $\mathcal{O}(H^2)$ terms. Throughout, we also don't have the $(1/\epsilon^2)$ suppression as $\epsilon=1$.

\section{Discussion}
\label{sec: discuss}
In this paper we have extended the Alcubierre warp drive spacetime to an infinite class of warp drives utilising the Martel--Poisson charts for Minkowski and AdS spacetimes and their conical defect cousins in three dimensions. The Minkowski spacetime admits Painlev\'{e}--Gull strand coordinates but for AdS spacetime the usual PG charts vanish and the warp drive spacetimes discussed in this paper are the only possibility. Though we have analysed the warp drives away from the conical singularities at the origin, the fact that spatial hypersurfaces are cones have global consequences which have been captured in the expressions for the expansion and $(-ve)$ energy densities localised around the warp drive boundary. To the leading order, energy requirements either scale up or down compared to the original Alcubierre warp drive depending upon the choice of the embedding. All the warp drives discussed in the paper violate the null energy condition (NEC). Some other physical attributes like the tilting of the light cones and formation of forward horizon are also affected by the conical defect backgrounds.

There are several possibilities for generalisations. We can look for warped spherically symmetric background spacetimes 
\begin{equation}
\label{eq: gen_metric}
ds^2=-e^{2\Phi} d t^2+\frac{d r^2}{f}+r^2 d \Omega_{(D-2)}^2
\end{equation}
which admits MP--charts with partial coverage of the spacetime, 
\begin{align}
\label{eq: gen_MP}
ds^2=& -dT^2 +\frac{e^{2\Phi}}{\epsilon^2 f}\left(dr +\frac{\epsilon\sqrt{f}}{e^\Phi}\sqrt{1-\frac{e^{2\Phi}}{\epsilon^2}} \vt dT\right)^2 \nonumber \\
&+r^2 d \Omega_{(D-2)}^2 \, , \quad \left(1-\frac{e^{2\Phi}}{\epsilon^2}\right)\geq 0 \, .
\end{align}
Another direction would be to look at higher dimensions $D\geq 4$ and find spacetimes whose MP--charts has spatial hypersurfaces which near the singularity at the origin takes the form, 
\begin{align}
\label{eq: cone_embed}
d s_C^2=& d \tilde{r}^2+\tilde{r}^2 d \tilde{\phi}^2 \\
& +\sum_{i, j=1}^{D-3}\left(w_{i j}(\theta)+h_{i j}(\theta) \tilde{r}^2\right) d \theta^i d \theta^j+\mathcal{O}(\tilde{r}^4) \nonumber
\end{align}
or generalise the definition of a cone from two to higher dimensions,
\begin{equation}
\label{eq: gen_cone}
ds_C^2= d\tilde{r}^2 +\tilde{r}^2 d \tilde{\Omega}_{(D-2)}^2 \, ,
\end{equation}
where unless $d \tilde{\Omega}_{(D-2)}^2$ is the metric of the unit sphere $S^{(D-2)}$, the generalised cone $ds_C^2$ is singular. For example, MP--charts of various black hole backgrounds would come under its purview. For most of these generalisations, the Ricci scalar of the spatial hypersurface will be non zero thereby opening up new possibilities for the energy density $\rho$. If we focus on the spherically symmetric cases, the expressions involving the extrinsic curvatures would generalise easily from the three dimensional expressions in the previous sections upto some coefficients. We have to symmetrise the expressions in all spatial coordinates except $x$, the direction of travel for the warp drive.  

While warp drives are beyond the realm of current experiments, various \textit{analog gravity} setups imitate the metric \cite{Fischer_2003, finazzi2012analogue} and can be studied experimentally \cite{PhysRevLett.105.240401}. It might be possible to generalize these setups to accommodate the Martel--Poisson versions of the warp drive. Various other properties of the wrap drive are of interest, for example the \textit{zero--expansion} drives of Nat\'{a}rio \cite{Nat_rio_2002}, the exact analysis of the formation of event horizon and various quantum effects  \cite{Hiscock_1997, Finazzi_2009, Barcel__2022},  optimisation of various parameters including the form functions and the idea of a warp bubble inside another bubble \cite{Broeck_1999, Helmerich_2023}. Another tangent line of investigation would be to explore the AdS/CFT dictionary and its consequences in the warp drive perturbed AdS metrics \cite{Maldacena:1997re}. We hope to study some of these aspects in our future works. 

\begin{acknowledgements}
A.C. thanks Manirujjaman Chowdhury for useful discussions . The work A.C. is supported by IIT Bhubaneswar Seed Grant SP–103.
\end{acknowledgements}

\begin{appendix}
\section{Different choices for \texorpdfstring{$r_s$}{} and \texorpdfstring{$\gamma_{ij}$}{}}
\label{app: r_s}
In this appendix we shall explore, two combinations of choices for the warp drive radial coordinate $r_s$ and the spatial induced metric $\gamma_{ij}$ as mentioned in section \ref{sec:2}. Our focus is on the warp drives embedded in the MP--charts of Minkowski spacetime but similar analysis goes through for the conical defect background, $\text{AdS}_3$ \& $\text{dS}_3$ spacetimes and their conical defects cousins.
\paragraph{\textbf{Case I :}}We take $r_s=\sqrt{(x-x_0)^2+y^2}$ and $\gamma_{ij}=\delta_{ij}+(1-f(r_s))\left(\frac{1}{\epsilon^2}-1\right)\frac{x^i x^j}{r^2}$. There are two important deviations from the expressions in \eqref{eq: wrap_mink}. Firstly, the effect of $(1-f(r_s))$ term in $\gamma_{ij}$ starts to show only at $\mathcal{O}(v_\infty^2)$, so if we are truncating to $\mathcal{O}(v_\infty)$, all results will match. Secondly, the spatial curvature tensors, particularly the Ricci scalar ${}^{(2)} R$ is not zero at finite $r$. Its contribution to the energy density $\rho$ starts at order $\mathcal{O}(v_\infty^2)$, 
\begin{equation}
\label{eq: appA_ricci}
\frac{(y^2-r_s^2)f'(r_s)-r_sy^2f''(r_s)}{r_s^3}v_\infty^2 + \mathcal{O}(\frac{1}{r})
\end{equation}
Similarly, the $K_{ij}$ in \eqref{eq: ext} now picks up contributions from the $\frac{1}{2} \partial_T \gamma_{ij}$ term, again modifying expressions from $\mathcal{O}(v_\infty^2)$ onwards. In all, this choice complicates things but after background subtraction, $\theta$ and $\rho$ are proportional to $f'(r_s)^{\#}$ and $f''(r_s)^{\#}$, therefore localised around the wrap drive bubble. A lot now depends on the choice of $f(r_s)$, tweaking it might give a $(+ve)$ ${}^{(2)}R$ thereby reducing the overall $(-ve)$ energy density required to operate the drive.
\paragraph{\textbf{Case II :}}We take $r_s=\sqrt{\gamma_{ij}(x-x_o(T))^i(x-x_0(T))^j}$ and $\gamma_{ij}=\delta_{ij}+\left(\frac{1}{\epsilon^2}-1\right)\frac{x^i x^j}{r^2}$. Here, as we are not changing the spatial metric, there is no contribution from ${}^{(2)}R$, neither $\frac{1}{2} \partial_T \gamma_{ij}$ contributes to $K_{ij}$. Extra contributions as compared to results in section \ref{sec:2} are due to the changes in the derivatives of $f(r_s)$ or equivalently $r_s$ w.r.t. $(x,\,y)$ coordinates. Again, the changes are from $\mathcal{O}(v_\infty^2)$ onwards and yet again the contributions to $\theta$ and $\rho$ are localised to the boundary of the warp drive.
\end{appendix}

\bibliographystyle{spphys}       
\bibliography{reference.bib}   

\begin{thebibliography}{10}
\providecommand{\url}[1]{{#1}}
\providecommand{\urlprefix}{URL }
\expandafter\ifx\csname urlstyle\endcsname\relax
  \providecommand{\doi}[1]{DOI \discretionary{}{}{}#1}\else
  \providecommand{\doi}{DOI \discretionary{}{}{}\begingroup
  \urlstyle{rm}\Url}\fi

\bibitem{10.1119/1.15620}
M.S. Morris, K.S. Thorne, American Journal of Physics \textbf{56}(5), 395
  (1988).
\newblock \doi{10.1119/1.15620}.
\newblock \urlprefix\url{https://doi.org/10.1119/1.15620}

\bibitem{Alcubierre_1994}
M.~Alcubierre, Classical and Quantum Gravity \textbf{11}(5), L73–L77 (1994).
\newblock \doi{10.1088/0264-9381/11/5/001}.
\newblock \urlprefix\url{http://dx.doi.org/10.1088/0264-9381/11/5/001}

\bibitem{Nat_rio_2002}
J.~Natário, Classical and Quantum Gravity \textbf{19}(6), 1157–1165 (2002).
\newblock \doi{10.1088/0264-9381/19/6/308}.
\newblock \urlprefix\url{http://dx.doi.org/10.1088/0264-9381/19/6/308}

\bibitem{Lobo_2004}
F.S.N. Lobo, M.~Visser, Classical and Quantum Gravity \textbf{21}(24),
  5871–5892 (2004).
\newblock \doi{10.1088/0264-9381/21/24/011}.
\newblock \urlprefix\url{http://dx.doi.org/10.1088/0264-9381/21/24/011}

\bibitem{Everett_1997}
A.E. Everett, T.A. Roman, Physical Review D \textbf{56}(4), 2100–2108 (1997).
\newblock \doi{10.1103/physrevd.56.2100}.
\newblock \urlprefix\url{http://dx.doi.org/10.1103/PhysRevD.56.2100}

\bibitem{Martel_2001}
K.~Martel, E.~Poisson, American Journal of Physics \textbf{69}(4), 476–480
  (2001).
\newblock \doi{10.1119/1.1336836}.
\newblock \urlprefix\url{http://dx.doi.org/10.1119/1.1336836}

\bibitem{Fischer_2003}
U.R. Fischer, M.~Visser, Europhysics Letters (EPL) \textbf{62}(1), 1–7
  (2003).
\newblock \doi{10.1209/epl/i2003-00103-6}.
\newblock \urlprefix\url{http://dx.doi.org/10.1209/epl/i2003-00103-6}

\bibitem{finazzi2012analogue}
S.~Finazzi.
\newblock Analogue gravitational phenomena in bose-einstein condensates (2012)

\bibitem{PhysRevLett.105.240401}
O.~Lahav, A.~Itah, A.~Blumkin, C.~Gordon, S.~Rinott, A.~Zayats, J.~Steinhauer,
  Phys. Rev. Lett. \textbf{105}, 240401 (2010).
\newblock \doi{10.1103/PhysRevLett.105.240401}.
\newblock
  \urlprefix\url{https://link.aps.org/doi/10.1103/PhysRevLett.105.240401}

\bibitem{Faraoni2020WhenPC}
V.~Faraoni, G.~Vachon, The European Physical Journal. C, Particles and Fields
  \textbf{80} (2020).
\newblock \urlprefix\url{https://api.semanticscholar.org/CorpusID:219956329}

\bibitem{garattini2023black}
R.~Garattini, K.~Zatrimaylov.
\newblock Black holes and warp drive (2023)

\bibitem{garattini2024wormholewarp}
R.~Garattini, K.~Zatrimaylov.
\newblock On the wormhole--warp drive correspondence (2024)

\bibitem{Fursaev_1995}
D.V. Fursaev, S.N. Solodukhin, Physical Review D \textbf{52}(4), 2133–2143
  (1995).
\newblock \doi{10.1103/physrevd.52.2133}.
\newblock \urlprefix\url{http://dx.doi.org/10.1103/PhysRevD.52.2133}

\bibitem{Bobrick_2021}
A.~Bobrick, G.~Martire, Classical and Quantum Gravity \textbf{38}(10), 105009
  (2021).
\newblock \doi{10.1088/1361-6382/abdf6e}.
\newblock \urlprefix\url{http://dx.doi.org/10.1088/1361-6382/abdf6e}

\bibitem{ellis2004warp}
H.G. Ellis.
\newblock The warp drive and antigravity (2004)

\bibitem{Broeck_1999}
C.V.D. Broeck, Classical and Quantum Gravity \textbf{16}(12), 3973–3979
  (1999).
\newblock \doi{10.1088/0264-9381/16/12/314}.
\newblock \urlprefix\url{http://dx.doi.org/10.1088/0264-9381/16/12/314}

\bibitem{Gourgoulhon:2012ffd}
E.~Gourgoulhon, \emph{{3+1 Formalism in General Relativity}}.
\newblock Lecture Notes in Physics (Springer, 2012).
\newblock \doi{10.1007/978-3-642-24525-1}

\bibitem{DESER1984220}
S.~Deser, R.~Jackiw, G.~{'t Hooft}, Annals of Physics \textbf{152}(1), 220
  (1984).
\newblock \doi{https://doi.org/10.1016/0003-4916(84)90085-X}.
\newblock
  \urlprefix\url{https://www.sciencedirect.com/science/article/pii/000349168490085X}

\bibitem{Kibble:1976sj}
T.W.B. Kibble, J. Phys. A \textbf{9}, 1387 (1976).
\newblock \doi{10.1088/0305-4470/9/8/029}

\bibitem{Randall_1999}
L.~Randall, R.~Sundrum, Physical Review Letters \textbf{83}(17), 3370–3373
  (1999).
\newblock \doi{10.1103/physrevlett.83.3370}.
\newblock \urlprefix\url{http://dx.doi.org/10.1103/PhysRevLett.83.3370}

\bibitem{Santiago_2022}
J.~Santiago, S.~Schuster, M.~Visser, Physical Review D \textbf{105}(6) (2022).
\newblock \doi{10.1103/physrevd.105.064038}.
\newblock \urlprefix\url{http://dx.doi.org/10.1103/PhysRevD.105.064038}

\bibitem{Kontou_2020}
E.A. Kontou, K.~Sanders, Classical and Quantum Gravity \textbf{37}(19), 193001
  (2020).
\newblock \doi{10.1088/1361-6382/ab8fcf}.
\newblock \urlprefix\url{http://dx.doi.org/10.1088/1361-6382/ab8fcf}

\bibitem{Hawking_Ellis_1973}
S.W. Hawking, G.F.R. Ellis, \emph{The Large Scale Structure of Space-Time}.
\newblock Cambridge Monographs on Mathematical Physics (Cambridge University
  Press, 1973)

\bibitem{Pfenning_1997}
M.J. Pfenning, L.H. Ford, Classical and Quantum Gravity \textbf{14}(7),
  1743–1751 (1997).
\newblock \doi{10.1088/0264-9381/14/7/011}.
\newblock \urlprefix\url{http://dx.doi.org/10.1088/0264-9381/14/7/011}

\bibitem{PhysRevD.43.3972}
L.H. Ford, Phys. Rev. D \textbf{43}, 3972 (1991).
\newblock \doi{10.1103/PhysRevD.43.3972}.
\newblock \urlprefix\url{https://link.aps.org/doi/10.1103/PhysRevD.43.3972}

\bibitem{Pfenning_1997_1}
M.J. Pfenning, L.H. Ford, Physical Review D \textbf{55}(8), 4813–4821 (1997).
\newblock \doi{10.1103/physrevd.55.4813}.
\newblock \urlprefix\url{http://dx.doi.org/10.1103/PhysRevD.55.4813}

\bibitem{Misner1964RELATIVISTICEF}
C.W. Misner, D.H. Sharp, Physical Review \textbf{136}, 571 (1964).
\newblock \urlprefix\url{https://api.semanticscholar.org/CorpusID:123043032}

\bibitem{Hernandez:1966zia}
W.C. Hernandez, C.W. Misner, Astrophys. J. \textbf{143}, 452 (1966).
\newblock \doi{10.1086/148525}

\bibitem{DESER1984405}
S.~Deser, R.~Jackiw, Annals of Physics \textbf{153}(2), 405 (1984).
\newblock \doi{https://doi.org/10.1016/0003-4916(84)90025-3}.
\newblock
  \urlprefix\url{https://www.sciencedirect.com/science/article/pii/0003491684900253}

\bibitem{Hiscock_1997}
W.A. Hiscock, Classical and Quantum Gravity \textbf{14}(11), L183 (1997).
\newblock \doi{10.1088/0264-9381/14/11/002}.
\newblock \urlprefix\url{https://dx.doi.org/10.1088/0264-9381/14/11/002}

\bibitem{Finazzi_2009}
S.~Finazzi, S.~Liberati, C.~Barceló, Physical Review D \textbf{79}(12) (2009).
\newblock \doi{10.1103/physrevd.79.124017}.
\newblock \urlprefix\url{http://dx.doi.org/10.1103/PhysRevD.79.124017}

\bibitem{Barcel__2022}
C.~Barceló, V.~Boyanov, L.J. Garay, E.~Martín-Martínez, J.M.
  Sánchez~Velázquez, Journal of High Energy Physics \textbf{2022}(8) (2022).
\newblock \doi{10.1007/jhep08(2022)288}.
\newblock \urlprefix\url{http://dx.doi.org/10.1007/JHEP08(2022)288}

\bibitem{Helmerich_2023}
C.~Helmerich, J.~Fuchs, A.~Bobrick, L.~Sellers, S.~Dangelo, G.~Martire, J.F.
  Agnew, in \emph{AIAA SCITECH 2023 Forum} (American Institute of Aeronautics
  and Astronautics, 2023).
\newblock \doi{10.2514/6.2023-0553}.
\newblock \urlprefix\url{http://dx.doi.org/10.2514/6.2023-0553}

\bibitem{Maldacena:1997re}
J.M. Maldacena, Adv. Theor. Math. Phys. \textbf{2}, 231 (1998).
\newblock \doi{10.4310/ATMP.1998.v2.n2.a1}

\end{thebibliography}

%
%


\end{document}